\RequirePackage{lineno}

\documentclass[prd,preprint,nofootinbib,superscriptaddress,floatfix]{revtex4-1}

\usepackage{cancel}
\usepackage{graphicx}
\usepackage{epsfig}
\usepackage{amsmath,amsfonts,amssymb}
\usepackage{t1enc}

\newcommand{\bY}{{\rm \bf Y}}
\newcommand{\bM}{{\rm \bf M}}
\newcommand{\bV}{{\rm \bf V}}
\newcommand{\bU}{{\rm \bf U}}
\newcommand{\phia}{\Phi_a}
\newcommand{\bA}{{\rm \bf A}}
\newcommand{\bB}{{\rm \bf B}}
\newcommand{\bX}{{\rm \bf X}}

\begin{document}

\title{\Large Radiative charged-lepton mass generation in\\ multi-Higgs doublet models}

\author{F.R. Joaquim}
\email{filipe.joaquim@tecnico.ulisboa.pt}
\author{J.T. Penedo}
\email{joao.t.n.penedo@tecnico.ulisboa.pt}
\affiliation{Departamento de F\'{\i}sica and CFTP, Instituto Superior T\'ecnico, Universidade de Lisboa, Lisboa, Portugal}

\begin{abstract}
We show that charged-lepton masses can be radiatively induced in multi-Higgs doublet models (NHDMs) through the renormalization group running of Yukawa couplings from high to low energies. Some extreme examples of electron and muon mass generation are discussed in the context of two and three Higgs doublet models. It is also shown that quantum corrections to the Yukawa couplings can be naturally of the same order as the tree-level values. We also comment on the implications of considering extensions of NHDMs in which right-handed neutrinos are added. 

\end{abstract}

\maketitle

\section{Introduction}

The discovery of a Standard Model (SM) Higgs-like boson at the Large Hadron Collider (LHC)~\cite{Aad:2012tfa} represents an important milestone on the understanding of mass generation and electroweak symmetry breaking (EWSB). Still, and in spite of decades of theoretical investigation, the reason why fermion masses span over several orders of magnitude remains in {\em terra incognita}. The quark and charged-lepton mass pattern suggests that third-generation masses arise at the classical level whereas the remaining ones originate from quantum corrections. This hypothesis, based on the possibility that the electron mass could be radiatively induced by the mass of the muon, was originally put forward by 't Hooft in his seminal work on the renormalizability of non-Abelian gauge theories~\cite{'tHooft:1971rn}. Attempts of implementing this idea in specific scenarios were pursued soon after by several authors~\cite{Weinberg:1972ws}. Later on, realistic models of radiative fermion-mass hierarchy were investigated in the framework of Grand Unified theories (GUTs). In such cases, light fermion masses are induced at the quantum level via the exchange of heavy particles with masses of the order of the GUT scale~\cite{Barr:1979xt}. The possibility of generating light fermion masses (and small mixing matrix elements) radiatively has also been investigated in supersymmetric theories~\cite{Nanopoulos:1982zm}.

An alternative explanation for fermion-mass hierarchies relies on imposing new symmetries which act on flavor space and, thus, constrain the structure of the Yukawa couplings. In this context, the most popular scenario is perhaps the one in which the shaping symmetries are Abelian U(1)'s broken by the vacuum expectation value (VEV) of some scalar {\em flavon} fields, $\langle S \rangle$. Below the typical mass scale $M$ of these scalars, effective Yukawa couplings are given by powers of small parameters $\epsilon \sim \langle S \rangle / M \ll 1$~\cite{Froggatt:1978nt}. More recently, a great deal of attention has been devoted to discrete symmetries in the explanation of neutrino masses and mixing~\cite{Altarelli:2010gt}. Some of these models postulate the existence of several Higgs doublets~\cite{Felipe:2013vwa} which couple to the fermion fields according to the rules imposed by broken or unbroken symmetries depending, for instance, on whether the Yukawa couplings are generated at the renormalizable level or not. The fermion mass pattern will then reflect the properties of the Yukawa couplings and also of the vacuum configuration. It is common practice to consider that a certain VEV pattern is not phenomenologically viable if it does not lead to the correct values of fermion masses at tree level. For instance, if at tree level a charged lepton $e_i$ is massless (or too light) in a multi-Higgs doublet model where only the neutral component of one of the doublets acquires a VEV, one would rule out the model. However, although not contributing classically to the mass $m_{e_i}$, Yukawa couplings between $e_i$ and  the zero-VEV Higgs doublets could induce, at the quantum level, a sizable coupling with the Higgs which acquires a VEV, thus generating (or contributing to) $m_{e_i}$ after EWSB.

In this Letter we explore the possibility of radiative charged-lepton mass generation due to the aforementioned corrections, considering also some simple examples in the 2HDM and 3HDM. To conclude, we briefly comment on the implications of our results in NHDMs extended with right-handed neutrinos.

\section{Radiative charged-lepton mass generation in NHDM\lowercase{s}}

Let us consider an extension of the SM with $N$ Higgs doublets $\phia=(\phi_a^+,\phi_a^0)^T$ transforming as $\phia \sim (2,1/2)$ under the SM gauge group SU(2)$_W\times$ U(1)$_Y$. The most general Yukawa Lagrangian for quarks and leptons is
\begin{equation}
\label{Lql}
-\mathcal{L}=(\bY_a^u)_{ij} \bar{q}_{Li} \tilde{\Phi}_a u_{Rj} +(\bY_a^d)_{ij} \bar{q}_{Li} \phia d_{Rj}  +(\bY_a^\ell)_{ij} \bar{\ell}_{Li} \phia e_{Rj} +{\rm H.c.}\,,
\end{equation}
where $q_{Li}$, $\ell_{Li}$ denote the quark and lepton doublets, while $u_{Ri}$, $d_{Ri}$ and $e_{Ri}$ are the quark and lepton right-handed singlets. As usual $\tilde{\Phi}_a=i\sigma_2 \phia^\ast=(\phi_a^{0\ast},-\phi_a^{-})^T$. The Yukawa matrices $\bY_a^X$ are general $3\times 3$ complex matrices diagonalized by biunitary transformations
\begin{equation}
\label{Biunit}
\bV_a^{X\dag} \bY_a^X \bU_a^X={\rm diag}(y_{a1}^X,y_{a2}^X,y_{a3}^X)\,,
\end{equation}
with $y_{ai}^X$ real and positive. After EWSB, tree-level mass matrices are generated as
\begin{equation}
\label{Ms}
\bM_{d,\ell}=\sum_{a=1}^N v_a \bY_a^{d,\ell}  \; , \;
\bM_{u}=\sum_{a=1}^N v_a^\ast \bY_a^u\,,
\end{equation}
where $v_a=\langle \phi_a^0 \rangle$ is the VEV of $\phi_a^0$. From now on, we restrict ourselves to the study of corrections to the charged-lepton Yukawa couplings. At the one-loop level, the renormalization group equations (RGEs) for $\bY_a^\ell$ read
\begin{equation}
\label{RGE}
16 \pi^2 \frac{d \bY_a^\ell}{dt}=\beta_{a}\;,\; t=\log\left(\frac{\mu}{\Lambda}\right)\,,
\end{equation}
where $\mu$ and $\Lambda$ are the renormalization and reference energy scales, respectively. For the NHDM, the corresponding beta function at one loop is~\cite{Grimus:2004yh} 
\begin{equation}
\label{beta}
\beta_{a}^{(1)}=\alpha_g \bY_a^\ell+\alpha_Y^{ab} \bY_b^\ell+\bY_a^\ell \bY_b^{\ell\dag} \bY_b^\ell+\frac{1}{2} \bY_b^\ell \bY_b^{\ell\dag} \bY_a^\ell\,,
\end{equation}
with $\alpha_g=-9g^2/4-15g^{\prime 2}/4$, being $g$ and $g^\prime$ the SU(2)$_W$ and U(1)$_Y$ gauge couplings, and 
\begin{equation}
\label{alphas}
\alpha_Y^{ab}=3{\rm Tr}(\bY_a^{u\dag} \bY_b^u+\bY_a^d\bY_b^{d\dag} )+{\rm Tr}(\bY_a^\ell \bY_b^{\ell\dag} )\,.
\end{equation}
In order to simplify our analysis, we henceforth consider the Higgs basis~\cite{Lavoura:1994fv} where all VEVs are zero except the one of some $\phi_a^0$ ($v_a=v=174$~GeV) and Yukawa couplings are rotated accordingly. Unless explicitly stated, we will always work in this basis (although our results hold for a general vacuum configuration). Therefore, the charged-lepton mass matrix is given by
\begin{equation}
\label{Mell}
\bM_\ell=v \bY_a^\ell\;,\; \bV_a^{\ell\dag} \bY_a^\ell \bU_a^\ell={\rm diag}(y_e,y_\mu,y_\tau)\,,
\end{equation}
at tree level. 

We are interested in studying the one-loop corrected Yukawa couplings, which we denote by $\bY_a^{\ell (1)}$. The beta function of $\bY_a^\ell$ given in eq. \eqref{beta} contains terms depending on $\bY_a^\ell$ (which contribute to the tree-level masses) and terms which are proportional to $\bY_b^\ell$, namely $\alpha_Y^{ab} \bY_b^\ell$. These stem from Higgs wave-function diagrams with ingoing $\Phi_a$, quarks and leptons in the loop, and outgoing $\Phi_b$. For $b\neq a$, such contributions induce corrections to $\bY_a^\ell$ which are independent from it. Under the reasonable assumption that the top quark couples with the same strength to all Higgs doublets with $y_t\simeq 1$, and considering all Yukawa matrices $\bY_b^\ell$ diagonal with elements given by $y_{bi}$, the corrections to $\bY_a^\ell$ coming from $\bY_b^\ell$ can typically be of the order of 
\begin{equation}
\label{deltay}
\delta y_{ai}\sim  \frac{3 y_{bi}}{16\pi^2}\log\left(\frac{\Lambda}{m_H}\right) \;,\; i=1,2,3\;,\; b\neq a\,,
\end{equation}
where $\Lambda$ is a high scale at which the Yukawa couplings are initially given, and $m_H$ is the typical scale of the extra scalars in the theory. This rough approximation illustrates the fact that the Yukawa couplings with the Higgses which do not acquire VEVs (and, therefore, do not contribute to masses at tree level) can induce important contributions to $\bY_a^\ell$, depending on the scales $\Lambda$ and $m_H$, and on the size of the couplings $\bY_b^\ell\,(b\neq a)$. As we will show later, one could even generate charged-lepton masses solely from these new contributions.

In order to have three massive charged leptons, we require $r(\bY_a^{\ell (1)})=3$, where $r(\bA)$ represents the rank of a matrix $\bA$. It is straightforward to check that quantum corrections in the context of the NHDM may affect the rank of the Yukawa coupling matrices, and thus induce lepton masses which are absent at tree-level. In particular, the beta function for $\bY_a^\ell$, at any loop order $n$, can be written in the form
\begin{equation}
\label{betaform}
\beta_a^{(n)} = \sum_{k=1}^N \, \bX_{a,k}^{(n)}\, \bY_k^\ell\, ,
\end{equation}
where $\bX_{a,k}^{(n)}$ are matrices in flavor space. For instance, from eq.~\eqref{beta}, one has
\begin{equation}
\label{X1loop}
\bX_{a,k}^{(1)} = \big(\delta_{ak}\,\alpha_g+\alpha_Y^{ak}\big)\,\openone + \bY_a^\ell {\bY_k^\ell}^\dagger + \frac{1}{2}\, \delta_{ak}\, \bY_a^\ell {\bY_a^\ell}^\dagger\,.
\end{equation}
Notice that the structure of $\bY_a^{\ell (n)}$ will be the same as that of eq.~\eqref{betaform} with different $\bX$ matrices, $\bX_{a,k}^{\prime(n)}$. Since $r(\bA \bB)\leq \min\{r(\bA),r(\bB)\}$, it is clear that the rank of each term in the sum of eq.~\eqref{betaform} is, at most, equal to $r(\bY_k^\ell)$.
Finally, as rank is subadditive, {\em i.e.} $r(\bA+\bB) \leq r(\bA)+ r(\bB)$, one has
\begin{equation}
\label{rankbeta}
r(\bY_a^{\ell (n)})=r\Bigg( \sum_{k=1}^N \, \bX_{a,k}^{\prime(n)}\, \bY_k^\ell\Bigg) \, \, \leq \, \, {\rm min}\left\{3,\sum_{k=1}^N \, r(\bY_k^\ell)\right\}\,.
\end{equation}
In fact, barring tailored cancellations among Yukawa structures, equality generally holds in the above equation. This result implies that one can in principle generate charged-lepton masses from low-rank tree-level Yukawa couplings. Hence, the number of massive charged leptons will depend on these ranks and on the number of existing Higgs doublets. For instance, to end up with $n_m$ massive charged leptons from rank-1 (tree-level) mass and Yukawa matrices one would need~\footnote{This condition is necessary but not sufficient due to the aforementioned possibility of cancellations. In fact, rank is not strictly additive if and only if the matrices column spaces or row spaces intersect (see for instance~\cite{DavidCallan:1998}). This happens in the trivial case $\bY_2^\ell \propto \bY_1^\ell$.} $N \geq n_m$. To make our point clear, let us consider a couple of very simple and extreme examples where radiative mass generation is important. Notice that it is not our goal to frame these examples in the context of specific models but to provide a proof of concept for our claims. Further investigation on this subject will be presented elsewhere~\cite{wip}. 

In the  2HDM case~\cite{Branco:2011iw}  only one charged-lepton mass can be radiatively generated from rank-1 Yukawa matrices in the Higgs basis. Instead, if $3-n$ charged leptons are massive at tree level, one needs $r(\bY_2^\ell) \geq n$ to end up with all charged leptons massive. In the 3HDM all three charged leptons could acquire mass from rank-1 Yukawa matrices (in the Higgs basis). To illustrate this, consider a 3HDM with a vacuum configuration of the type $(v_1,v_2,v_3)=(0,0,v)$ and with Yukawa couplings~\footnote{Notice that in the following examples the zero entries in the Yukawa matrices should perhaps not be taken as strict zeros but interpreted as the limit of having very suppressed entries. We also remark that our results are valid for more complicated flavor structures which we do not consider here.} $\bY_1^\ell={\rm diag}(\epsilon_1,0,0)$, $\bY_2^\ell={\rm diag}(0,\epsilon_2,0)$, $\bY_3^\ell={\rm diag}(0,0,\epsilon_3)$. For simplicity, we consider all parameters real neglecting possible CP-violating effects in the leptonic sector~\cite{Branco:2011zb}. Within this setup, it is straightforward to see that only the tau is massive at tree level. Taking into account the beta function given in eq.~(\ref{beta}) and the fact that, due to the vacuum configuration, only  $\bY_3^\ell$ is relevant for the charged-lepton masses, we obtain
\begin{equation}
\label{Mlex}
m_{e_{i}}^{(1)}\simeq \frac{v}{16 \pi^2}\alpha_Y^{3i} \epsilon_i \log\left( \frac{\Lambda}{m_H}\right)\;,\; i=1,2\,,
\end{equation}
at one loop in the leading-log approximation. Notice that we are not interested here in possible flavor changing neutral current constraints~\cite{Mahmoudi:2009zx} since these can be avoided considering the decoupling limit of NHDMs~\cite{Haber:1989xc}. In our examples we will assume $\Lambda = \Lambda_{\rm GUT} \sim 10^{16}$~GeV and $m_H \sim 1$~TeV. Roughly taking  $\alpha_Y^{3i} \sim \mathcal{O}(1)$ we would get the right charged-lepton masses for $(\epsilon_1,\epsilon_2,\epsilon_3)\simeq(1.55\times 10^{-5},3.20\times 10^{-3},0.01)$. The hierarchy in the $\epsilon_i$'s could be possibly explained imposing U(1) symmetries {\em \`{a} la} Froggatt-Nielsen~\cite{Froggatt:1978nt}. If the structure of the couplings $\bY_a^\ell$ were non-diagonal, not only masses would be generated but also corrections to lepton mixing would arise. Although a general treatment of how mixing is corrected in the NHDM can be carried out, we just present a simple example which illustrates the effect.  Consider the same 3HDM as before with the following rank-1 Yukawa matrices,
\begin{align}
\label{Ysexam}
&\bY_1^\ell=
\left(
\begin{array}{ccc}
\epsilon_1 & 0 &0 \\
- \epsilon_1 \epsilon/\sqrt{2}  &0 &0 \\
- \epsilon_1\epsilon /\sqrt{2}  &0 &0 \\
\end{array}
\right)\;,\; %
\bY_2^\ell=\left(
\begin{array}{ccc}
0 &\epsilon_2\epsilon /\sqrt{2}   & 0  \\
0 & \epsilon_2 &0 \\
0 &0 &0
\end{array}
\right)\;,\;
\bY_3^\ell=\left(
\begin{array}{ccc}
0 &0  & \epsilon_3 \epsilon/\sqrt{2}   \\
0 & 0 & 0 \\
0 &0 &\epsilon_3
\end{array}
\right)\,.
\end{align}
Due to the $(0,0,v)$ vacuum, only one charged lepton is massive at tree level. Taking radiative effects into account, the Yukawa matrix $\bY_3^\ell$ would be corrected to $\bY_3^{\ell (1)}=\bY_3^{\ell}-\delta\bY_3^{\ell}$, where
\begin{align}
\label{Y3ex}
\delta\bY_3^{\ell}\simeq \frac{1}{16\pi^2}
\left(
\begin{array}{ccc}
\alpha_Y^{31}\epsilon_1 & \dfrac{\alpha_Y^{32} \epsilon_2\epsilon}{\sqrt{2}}  &\dfrac{\epsilon_3 \epsilon(\epsilon_2^2 \epsilon^2+6\epsilon_3^2)}{4\sqrt{2}} \medskip\\
- \dfrac{\alpha_Y^{31}\epsilon_1 \epsilon}{\sqrt{2}}  &\alpha_Y^{32}\epsilon_2 &\dfrac{1}{4}\epsilon_2^2\epsilon_3\epsilon^2 \medskip\\
- \dfrac{\alpha_Y^{31}\epsilon_1 \epsilon}{\sqrt{2}}   & 0 &\dfrac{3}{2}\epsilon_3^3\\
\end{array}\right)\log\left(\dfrac{\Lambda}{m_H}\right)\,,
\end{align}
with $\alpha_Y^{3i}$ given in eq.~(\ref{alphas}). Recalling that the mass matrix is $\bM_\ell^{(1)} = v \bY_3^{\ell (1)}$, the following masses are obtained
\begin{align}
\label{msexa}
m_{e}^{(1)}\simeq \frac{\alpha_Y^{31}\epsilon_1}{16\pi^2} \sqrt{1+\epsilon^2}\, v \log\left(\frac{\Lambda}{m_H}\right) \;,\;
m_{\mu}^{(1)}\simeq \frac{\alpha_Y^{32}\epsilon_2 }{16\pi^2} \sqrt{\frac{2+\epsilon^2}{2}}\, v\log\left(\frac{\Lambda}{m_H}\right)\,,
%
%
\end{align}
at the one-loop level. Taking $(\epsilon_1,\epsilon_2,\epsilon_3)\simeq (6.6\times 10^{-4},0.14,0.01)$, $\alpha_Y^{3i}\sim \mathcal{O}(1)$ and $\epsilon \ll 1$, these expressions reproduce the observed values for $m_{e,\mu,\tau}$. The difference between this example and the previous one is that the flavor structure of the charged-lepton mass matrix is no longer trivial. In fact, it can be shown that the left-handed rotation which brings $\bM_\ell^{(1)}$ to the diagonal basis is
\begin{equation}
\bV_L\simeq\left(
\begin{array}{ccc}
-1+\dfrac{\epsilon^2}{2} &\dfrac{\epsilon}{\sqrt{2}}   &\dfrac{\epsilon}{\sqrt{2}}  \\
\dfrac{\epsilon}{\sqrt{2}} &1- \dfrac{\epsilon^2}{4}&0 \\
\dfrac{\epsilon}{\sqrt{2}}  &-\dfrac{\epsilon^2}{2}  &1- \dfrac{\epsilon^2}{4}
\end{array}
\right)\,.
\end{equation}
This rotation would be extremely important if, for instance, the neutrino mass matrix exhibited a tribimaximal (TBM) mixing pattern~\cite{Harrison:2002er} (which is now experimentally excluded due to the fact that the reactor neutrino angle is nonzero). If this were the case in our example, then the lepton mixing matrix would have to be corrected from $\bU_{\rm TBM}$ to $\bV_L^\dag\bU_{\rm TBM}$, in order to account for the transformation to the diagonal basis of the charged-lepton left-handed fields. This would result in the following mixing angles
\begin{align}
\label{mixang}
\sin^2\theta_{12} \simeq \frac{1}{3}+\frac{2\sqrt{2}}{9}\epsilon\;,
\;\sin^2\theta_{23} \simeq \frac{1}{2}-\frac{\epsilon^2}{4}\;,\;
\sin^2\theta_{13}\simeq  \epsilon^2\,,
\end{align}
which, for $|\epsilon| \simeq 0.15$, lead to the right value for the reactor neutrino angle $\theta_{13}$, keeping the remaining mixing angles within their experimentally allowed ranges~\cite{Tortola:2012te}. This simple example shows how a scenario which would be excluded by tree-level considerations becomes phenomenologically viable when quantum corrections to the charged-lepton Yukawa couplings are included. We point out that we have started from a situation where the muon and the electron were massless and $\theta_{13}=0$, to end up with a case where $m_e$, $m_\mu$ and $\theta_{13}$ are radiatively generated.

Although we have presented examples with massless charged leptons at tree level -- which would certainly call for a justification -- we stress that, even if this is not the case, the RGE corrections to the Yukawas in NHDMs should always be kept in mind. To illustrate this, consider a 2HDM with real and diagonal Yukawa matrices $\bY_a^\ell={\rm diag}(y_{a1},y_{a2},y_{a3})$ in the Higgs basis where $\langle \phi_2^0 \rangle=0$. The one-loop corrected Yukawas will be
\begin{equation}
\label{yis}
y_{1i}^{(1)} \simeq y_{1i} - \frac{(\delta_{1b}\alpha_g+\alpha_Y^{1b})}{16 \pi^2}y_{bi} \log\left(\dfrac{\Lambda}{m_H}\right)\,.
\end{equation}
Taking the natural value $\alpha_Y^{12} \sim \mathcal{O}(1)$, it is apparent from the above estimate that if $y_{2i} \gtrsim 16\pi^2 y_{1i}/ \log(\Lambda/m_H)$, the one-loop contributions to the Yukawa couplings will always be important. For instance, if the electron couples with $\Phi_1$ and $\Phi_2$ with $y_{1e} = m_e/v \simeq 2.9 \times 10^{-6} $ and $y_{2e} \gtrsim 4.6\times 10^{-4}/\log(\Lambda/m_H)$, the correction to the electron Yukawa coupling is of the same order (or larger) than the tree-level value. If $\Lambda\sim 10^{16}$~GeV  and $m_H \sim 1$~TeV, this would require $y_{2e} \gtrsim 3.5\times 10^{-5}$, which can still be a very small number. Following the same argument for the tau lepton, $y_{2\tau} \gtrsim 0.1$ would lead to a correction of the order of $y_{1\tau}=m_\tau/v$. In fact, even if each charged-lepton couples with the same strength with the two Higgs doublets, the Yukawa couplings would become $y_{1i}^{(1)}=y_{1i} -\delta y_{1i}$, where
\begin{equation}
\label{yisratio}
\frac{\delta y_{1i}}{y_{1i}}\simeq  \frac{(\alpha_g+\alpha_Y^{11}+\alpha_{Y}^{12})}{16 \pi^2}\log\left(\dfrac{\Lambda}{m_H}\right)\,.
\end{equation}
With $\alpha_g+\alpha_Y^{11}+\alpha_{Y}^{12} \simeq  4$ (see eq.~\eqref{alphas}), $\Lambda\sim 10^{16}$~GeV  and $m_H \sim 1$~TeV, the induced $\delta y_{1i}$ amounts to approximately $70\%$ of the tree-level masses.

We now move into a brief digression on NHDMs extended with three right-handed neutrinos $\nu_{Rj}$ of masses $M_j \gg v$.   
In this case, the Yukawa Lagrangian reads 
\begin{equation}
\label{Lqlnu}
\mathcal{L}^\prime=\mathcal{L}-\left[(\bY_a^\nu)_{ij} \bar{\ell}_{Li} \tilde{\Phi}_a \nu_{Rj} +{\rm H.c.}\right]\,,
\end{equation}
where $\mathcal{L}$ has been given in eq.~(\ref{Lql}), and $\bY_a^\nu$ are the Dirac neutrino Yukawa coupling matrices. 
In the SM, the presence of couplings $\bY_\nu$ cannot generate charged-lepton masses radiatively since the new terms in the beta function of $\bY_\ell$ are of the form $\bY_\nu \bY_\nu^\dag \bY_\ell$. Thus, the charged-lepton Yukawa eigenvalues will always be proportional to themselves~\cite{Antusch:2005gp}. In contrast, the $\bY_k^\nu$ in the NHDM will contribute to the renormalization of $\bY_j^\ell$ both through wave-function and vertex corrections above the right-handed neutrino mass scale. Following the rules given in Refs.~\cite{Cheng:1973nv} for the calculation of RGEs in general gauge theories, one can show that the new beta function $\beta_{a}^\prime$, valid from the scale $\Lambda > M_i$ to $M_i$,  is at one loop
\begin{equation}
\label{betanu}
\beta_{a}^{\prime(1)}=\beta_{a}^{(1)}+\alpha_\nu^{ab} \bY_b^\ell+\frac{1}{2} \bY_b^\nu \bY_b^{\nu\dag} \bY_a^\ell   -2\bY_b^\nu\bY_a^{\nu\dag}\bY_b^\ell\,,
\end{equation}
with $\alpha_\nu^{ab}={\rm Tr}(\bY_a^{\nu\dag} \bY_b^\nu)$. Our result agrees with that presented in Ref.~\cite{Ibarra:2011gn}. Notice that the right-handed neutrinos add two new contributions to the beta function $\beta_a^{(1)}$ which do not depend on $\bY_a^\ell$ (cf.~eqs.~\eqref{beta}~and~\eqref{betanu}). However, these terms are only active from $\Lambda$ to $M_i$. 

Since seesaw light neutrino masses $m_\nu \sim 0.05\,{\rm eV}$ require $M_i\sim 10^{14}$~GeV for $\bY_k^\nu\sim \mathcal{O}(1)$, we can easily see that the corrections to the charged-lepton masses due to the $\bY_k^\nu$'s are at most $10\%$.

\section{Conclusions}

In this Letter we have shown that, in NHDMs, charged-lepton masses can be radiatively generated in a natural way due to the presence of Yukawa couplings of leptons and quarks with the extra non-SM Higgs doublets. We stress that these corrections should be taken into account in phenomenological studies of fermion mass and mixing models with more than one Higgs doublet. This could, for instance, be the case of some neutrino mass and mixing scenarios based on discrete symmetries. Hence, corrections to tree-level eigenvalues induced by the Yukawa couplings of all Higgs doublets should generally be considered in this class of models, especially if one aims at comparing model predictions with experiment. It is also important to mention that, besides affecting charged-lepton masses, the RGE running may have dramatic effects on the left-handed rotation which brings the charged-lepton mass matrix to its diagonal form~\cite{Ibarra:2011gn}. This will obviously alter the predictions for lepton mixing parameters, which have to be confronted with neutrino oscillation data. More detailed studies on this subject and its impact on fermion mass models with more than one Higgs doublet will be presented elsewhere~\cite{wip}.
\vspace*{5mm}

{\bf Note added:} While this Letter was being finalized, a related work appeared~\cite{Ibarra:2014fla},  where it is shown that quark masses and mixing can also be generated radiatively through the same effect discussed here.
\acknowledgments
We thank L. Lavoura and J.P. Silva for discussions. This work has been supported by the projects EXPL/FIS-NUC/0460/2013, CERN/FP/123580/2011 and PEst-OE-FIS-UI0777-2013, financed by \textit{Funda\c c\~ao para a Ci\^encia e a Tecnologia} (FCT, Portugal).

\end{document}